\documentclass[times, twocolumn]{aastex631_arxiv}
\usepackage{CJK}
\usepackage{graphicx}
\usepackage{enumerate}
\usepackage{amssymb, amsmath}
\usepackage{color}
\usepackage{ulem}
\usepackage{appendix}
\usepackage{comment}

\usepackage[nodayofweek]{datetime}
\newdateformat{monthyearday}{%
  \THEYEAR\ \monthname[\THEMONTH] \twodigit{\THEDAY}}

\submitjournal{AAS Journals}
\shorttitle{Keck/NIRC2 discovery of J1446B}
\shortauthors{Uyama et al.}

\begin{document}

\title{Direct Imaging Explorations for Companions from the Subaru/IRD Strategic Program II; Discovery of a Brown-dwarf Companion around a Nearby mid-M~dwarf LSPM~J1446+4633}

\correspondingauthor{Taichi Uyama}
\email{taichi.uyama.astro@gmail.com}

\author[0000-0002-6879-3030]{Taichi Uyama}
    \affiliation{Department of Physics and Astronomy, California State University Northridge, 18111 Nordhoff Street, Northridge, CA 91330, USA}
\author[0000-0002-4677-9182]{Masayuki Kuzuhara}
    \affiliation{Astrobiology Center, 2-21-1 Osawa, Mitaka, Tokyo 181-8588, Japan}
    \affiliation{National Astronomical Observatory of Japan, 2-21-1 Osawa, Mitaka, Tokyo 181-8588, Japan}
\author[0000-0002-5627-5471]{Charles Beichman}
    \affiliation{NASA Exoplanet Science Institute, 1200 E. California Blvd., Pasadena, CA 91125, USA}
    \affiliation{Infrared Processing and Analysis Center, California Institute of Technology, 1200 E. California Blvd., Pasadena, CA 91125, USA}
\author[0000-0003-3618-7535]{Teruyuki Hirano}
    \affiliation{Astrobiology Center, 2-21-1 Osawa, Mitaka, Tokyo 181-8588, Japan}
\author[0000-0001-6181-3142]{Takayuki Kotani}
    \affiliation{Astrobiology Center, 2-21-1 Osawa, Mitaka, Tokyo 181-8588, Japan}
    \affiliation{National Astronomical Observatory of Japan, 2-21-1 Osawa, Mitaka, Tokyo 181-8588, Japan}
    \affiliation{Department of Astronomical Science, The Graduate University for Advanced Studies, SOKENDAI, 2-21-1Osawa, Mitaka, Tokyo 181-8588, Japan}
\author[0000-0003-0115-547X]{Qier An}
    \affiliation{Department of Physics and Astronomy Johns Hopkins University, Baltimore, MD 21218, USA}
\author[0000-0003-2630-8073]{Timothy D. Brandt}
    \affiliation{Space Telescope Science Institute, 3700 San Martin Drive, Baltimore, MD 21218, USA}
    
\author[0000-0001-8345-593X]{Markus Janson}
    \affiliation{Department of Astronomy, Stockholm University, AlbaNova University Center, SE-10691, Stockholm, Sweden}
\author[0000-0002-8895-4735]{Dimitri Mawet}
    \affiliation{Department of Astronomy, California Institute of Technology, 1200 E. California Blvd., Pasadena, CA 91125, USA}
    \affiliation{Jet Propulsion Laboratory, California Institute of Technology, 4800 Oak Grove Dr., Pasadena, CA 91109, USA}
\author[0000-0003-1368-6593]{Mayuko Mori}
    \affiliation{Astrobiology Center, 2-21-1 Osawa, Mitaka, Tokyo 181-8588, Japan}
\author[0000-0001-8033-5633]{Bun'ei Sato}
    \affiliation{Department of Earth and Planetary Sciences, School of Science, Institute of Science Tokyo, 2-12-1 Ookayama, Meguro, Tokyo 152-8551, Japan}
\author[0000-0001-6277-9644]{Denitza Stoeva}
    \affiliation{Department of Astronomy, Faculty of Physics, Sofia University “St Kliment Ohridski”, 5 James Bourchier Blvd, 1164 Sofia, Bulgaria}
\author[0000-0002-6510-0681]{Motohide Tamura}
    \affiliation{Astrobiology Center, 2-21-1 Osawa, Mitaka, Tokyo 181-8588, Japan}

\author[0000-0001-8877-4497]{Masataka Aizawa}
    \affiliation{College of Science, Ibaraki University, 2-1-1 Bunkyo, Mito, 310-8512, Ibaraki, Japan}
\author[0000-0001-6279-0595]{Bryson Cale}
    \affiliation{Jet Propulsion Laboratory, California Institute of Technology, 4800 Oak Grove Dr., Pasadena, CA 91109, USA}
\author[0000-0002-1493-300X]{Thomas Henning}
    \affiliation{Max-Planck-Institut f\"{u}r Astronomie, K\"{o}nigstuhl 17, 69117 Heidelberg, Germany}
\author[0000-0001-6309-4380]{Hiroyuki Tako Ishikawa}
    \affiliation{Department of Physics and Astronomy, The University of Western Ontario, 1151 Richmond St, London, Ontario, N6A 3K7, Canada}
\author[0000-0001-8511-2981]{Norio Narita}
    \affiliation{Komaba Institute for Science, The University of Tokyo, 3-8-1 Komaba, Meguro, Tokyo 153-8902, Japan}
    \affiliation{Astrobiology Center, 2-21-1 Osawa, Mitaka, Tokyo 181-8588, Japan}
    \affiliation{Instituto de Astrof\'{i}sica de Canarias, 38205 La Laguna, Tenerife, Spain}
\author[0000-0002-8300-7990]{Masahiro Ogihara}
    \affiliation{Tsung-Dao Lee Institute, Shanghai Jiao Tong University, 1 Lisuo Road, Shanghai 201210, China}
    \affiliation{School of Physics and Astronomy, Shanghai Jiao Tong University, 800 Dongchuan Road, Shanghai 200240, China}
\author[0000-0002-1838-4757]{Aniket Sanghi}
    \altaffiliation{NSF Graduate Research Fellow}
    \affiliation{Department of Astronomy, California Institute of Technology, 1200 E. California Blvd., Pasadena, CA 91125, USA}
\author[0000-0002-0236-775X]{Trifon Trifonov}
    \affiliation{Max-Planck-Institut f\"{u}r Astronomie, K\"{o}nigstuhl 17, 69117 Heidelberg, Germany}
    \affiliation{Department of Astronomy, Faculty of Physics, Sofia University “St Kliment Ohridski”, 5 James Bourchier Blvd, 1164 Sofia, Bulgaria}
\author[0000-0002-6618-1137]{Jerry Xuan}
    \affiliation{Department of Astronomy, California Institute of Technology, 1200 E. California Blvd., Pasadena, CA 91125, USA}

\author[0000-0002-5082-8880]{Eiji Akiyama}
    \affiliation{Field of Fundamental Education and Liberal Arts, Department of Engineering, Niigata Institute of Technology, 1719 Fujihashi, Kashiwazaki, Niigata, 945-1195, Japan}
\author[0000-0002-7972-0216]{Hiroki Harakawa}
    \affiliation{Subaru Telescope, National Astronomical Observatory of Japan, National Institutes of Natural Sciences, 650 North A`oh$\bar{o}$k$\bar{u}$ Place, Hilo, HI 96720, USA}
\author{Klaus Hodapp}
    \affiliation{University of Hawaii, Institute for Astronomy, 640 N. Aohoku Place, Hilo, HI 96720, USA}
\author[0000-0003-1906-4525]{Masato Ishizuka}
    \affiliation{Department of Astronomy, The University of Tokyo, 7-3-1, Hongo, Bunkyo-ku, Tokyo 113-0033, Japan}
\author{Shane Jacobson}
    \affiliation{University of Hawaii, Institute for Astronomy, 640 N. Aohoku Place, Hilo, HI 96720, USA}
\author[0000-0002-5486-7828]{Eiichiro Kokubo}
    \affiliation{Center for Computational Astrophysics, National Astronomical Observatory of Japan, 2-21-1 Osawa, Mitaka, Tokyo 181-8588, Japan}
\author{Mihoko Konishi}
    \affiliation{Faculty of Science and Technology, Oita University, 700 Dannoharu, Oita 870-1192, Japan}
\author[0000-0002-9294-1793]{Tomoyuki Kudo}
    \affiliation{Subaru Telescope, National Astronomical Observatory of Japan, National Institutes of Natural Sciences, 650 North A`oh$\bar{o}$k$\bar{u}$ Place, Hilo, HI 96720, USA}
\author{Takashi Kurokawa}
    \affiliation{Astrobiology Center, 2-21-1 Osawa, Mitaka, Tokyo 181-8588, Japan}
    \affiliation{Institute of Engineering, Tokyo University of Agriculture and Technology, 2-24-16 Nakacho, Koganei, Tokyo 184-8588, Japan}
\author[0000-0003-2815-7774]{Jungmi Kwon}
    \affiliation{Department of Astronomy, Graduate School of Science, The University of Tokyo, 7-3-1 Hongo, Bunkyo-ku, Tokyo 113-0033, Japan}
\author[0000-0001-9326-8134]{Jun Nishikawa}
    \affiliation{National Astronomical Observatory of Japan, 2-21-1 Osawa, Mitaka, Tokyo 181-8588, Japan}
    \affiliation{Astrobiology Center, 2-21-1 Osawa, Mitaka, Tokyo 181-8588, Japan}
    \affiliation{Department of Astronomical Science, The Graduate University for Advanced Studies, SOKENDAI, 2-21-1Osawa, Mitaka, Tokyo 181-8588, Japan}
\author[0000-0002-5051-6027]{Masashi Omiya}
    \affiliation{Astrobiology Center, 2-21-1 Osawa, Mitaka, Tokyo 181-8588, Japan}
    \affiliation{National Astronomical Observatory of Japan, 2-21-1 Osawa, Mitaka, Tokyo 181-8588, Japan}
\author{Takuma Serizawa}
    \affiliation{Institute of Engineering, Tokyo University of Agriculture and Technology, 2-24-16 Nakacho, Koganei, Tokyo 184-8588, Japan}
    \affiliation{National Astronomical Observatory of Japan, 2-21-1 Osawa, Mitaka, Tokyo 181-8588, Japan}
\author[0000-0003-3860-6297]{Huan-Yu Teng}
    \altaffiliation{East Asian Core Observatories Association fellow}
    \affiliation{CAS Key Laboratory of Optical Astronomy, National Astronomical Observatories, Chinese Academy of Sciences, Beijing 100012, China}
    \affiliation{Department of Earth and Planetary Sciences, School of Science, Institute of Science Tokyo, 2-12-1 Ookayama, Meguro, Tokyo 152-8551, Japan}
\author{Akitoshi Ueda}
    \affiliation{Astrobiology Center, 2-21-1 Osawa, Mitaka, Tokyo 181-8588, Japan}
    \affiliation{National Astronomical Observatory of Japan, 2-21-1 Osawa, Mitaka, Tokyo 181-8588, Japan}
    \affiliation{Department of Astronomical Science, The Graduate University for Advanced Studies, SOKENDAI, 2-21-1Osawa, Mitaka, Tokyo 181-8588, Japan}
\author[0000-0003-4018-2569]{Sebastien Vievard}
    \affiliation{Subaru Telescope, National Astronomical Observatory of Japan, National Institutes of Natural Sciences, 650 North A`oh$\bar{o}$k$\bar{u}$ Place, Hilo, HI 96720, USA}


\begin{abstract}

We report the discovery of a new directly-imaged brown dwarf companion with Keck/NIRC2+pyWFS around a nearby mid-type M~dwarf LSPM~J1446+4633 (hereafter J1446). The $L'$-band contrast ($4.5\times10^{-3}$) is consistent with a $\sim20-60\ M_{\rm Jup}$ object at 1--10~Gyr and our two-epoch NIRC2 data suggest a $\sim30\%$ ($\sim3.1\sigma)$ variability in its $L'$-band flux. We incorporated Gaia DR3 non-single-star catalog into the orbital fitting by combining the Subaru/IRD RV monitoring results, NIRC2 direct imaging results, and Gaia proper motion acceleration. As a result, we derive ${59.8}_{-1.4}^{+1.5}\ M_{\rm Jup}$ and $\approx4.3~{\rm au}$ for the dynamical mass and the semi-major axis of the companion J1446B, respectively. 
J1446B is one of the intriguing late-T~dwarfs showing variability at $L'$-band for future atmospheric studies with the constrained dynamical mass.
Because the J1446 system is accessible with various observation techniques such as astrometry, direct imaging, and high-resolution spectroscopy including radial velocity measurement, it has a potential as a great benchmark system to improve our understanding for cool dwarfs. 

\end{abstract}


\section{Introduction} \label{sec: Introduction}
M~dwarfs are the most common stars in our galaxy \citep[e.g.][]{Reid1997,Reid2004,Winters2015} and have been the subject of numerous astronomical studies. Stellar multiplicity is one of the main topics for the M~dwarf studies related to star/planet formation and evolution mechanisms because a significant fraction of stellar systems are binaries or higher-order multiples \citep[e.g.][]{Janson2012,Janson2014,Winters2019}. However, multiplicity census surveys have been observationally biased because of unresolved binaries that could leave systematics and uncertainties of multiplicity. 
In addition, brighter M dwarfs have been targeted so far, limiting the census to earlier-type M dwarfs (see below). 
Recently, precise {\it Gaia} astrometry \citep{Gaia2016} has provided more observational information related to these studies.
Combination of astrometry with precise radial velocity (RV) surveys \citep[e.g., the Calar Alto high-Resolution search for M~dwarfs (CARMENES);][]{Cifuentes2025} or speckle imaging \citep[e.g.,][]{Clark2022} revealed more companions, which suggested that the multiplicity fraction of M~dwarfs could be underestimated. 

Particularly, substellar-mass companions ($\lesssim75\ M_{\rm Jup}$) around M~dwarfs within a few tens of au (corresponding to a few arcseconds for nearby M~dwarfs) have not well been explored; most of the nearby M~dwarfs are too faint for Hipparcos whose limiting magnitude is $V\sim12-13\ {\rm mag}$, so predicting unseen companions via Gaia-Hipparcos proper motion acceleration \citep{Brandt2018} is unavailable. Using only RV leaves a degeneracy between mass and inclination, making it impossible to put a strong constraint on mass. Spatially resolving the companion from the central star by high-contrast imaging is critical to solve this degeneracy. From ground, adaptive optics is essential for high-contrast imaging, and wavefront sensing is the key technique to achieve wavefront corrections. Classically only optical wavelengths were used for wavefront sensing and faint mid-late M~dwarfs ($R\gtrsim14$~mag) have not been prioritized as high-contrast imaging targets \citep[e.g.,][]{Bowler2015,Sanghi2024}. 
Instrumental developments, particularly near-infrared (NIR) pyramid wavefront sensing \citep{Peter2008,Bond2020,Lozi2022}, enabled high-contrast imaging observations of such faint M~dwarfs.
\cite{Uyama2023} conducted follow-up Keck/NIRC2+pyWFS observations targeting nearby mid--late M~dwarfs with signatures of long-term RV trends, based on the strategic program with the InfraRed Doppler (IRD) NIR high-dispersion spectroscopy at the Subaru telescope \citep[$R\sim70000$ covering $0.97-1.75\ \mu$m;][]{Kotani2018} monitoring RV perturbations of nearby mid-late M~dwarfs (hereafter IRD-SSP; 175 nights were assigned between 2019 and 2025; PI: Bun'ei Sato). These NIRC2 observations discovered new companions at projected separations of 2--20~au and demonstrated the feasibility to detect brown-dwarf (BD) companions within such separations. 

LSPM~J1446+4633 (hereafter J1446) is one of the nearby mid-M~dwarfs \citep[$T_{\rm eff}\sim3150\ {\rm K}$, SpT $\sim$~M4.5V, {\bf $M_{\rm star}\sim0.155\ M_\odot$};][]{Dittmann2014,TIC,Hejazi2022} at a distance of $\sim17\ {\rm pc}$ \citep{GaiaDR3}. The IRD-SSP observations detected significant RV perturbations and Renormalised Unit Weight Error (RUWE) from Gaia DR3 is 9.285, suggesting a massive companion around J1446. We conducted Keck/NIRC2+pyWFS high-contrast imaging as a continued high-contrast follow-up program of \cite{Uyama2023}. Section~\ref{sec: Data} briefly describes our Subaru/IRD RV monitoring and Keck/NIRC2 high-contrast imaging observations. We present each of the observational results in Section~\ref{sec: Results} with the discovery of the companion J1446B, and discuss the nature of the J1446 system in Section~\ref{sec: Discussions}.

\section{Data} \label{sec: Data}

We followed the same strategy as \cite{Uyama2023} for observations and data reduction of both Subaru/IRD and Keck/NIRC2 data sets (see Section~2.1 and 2.2 in the reference for details of IRD and NIRC2 data reduction respectively).

\subsection{Observations}

For the RV data, we obtained spectra of J1446 with Subaru/IRD 14 times as part of the IRD-SSP campaign between 2019 April and 2025 March. We utilized the data reduction pipeline \citep{Hirano2020} to extract the spectra and calculated RV variations using the spectra of the laser frequency comb deployed in Subaru/IRD as wavelength-stability references. 
We note that the two post-processing methods adopted in \citet{Kuzuhara2024} and \citet{Gorrini_2023_CARMENES}  were not applied to correct the RV fluctuations, because the first process is less suitable for the small RV sample size and the host star of a significant RV variation like J1446 and the second is unnecessary for tracking such a large RV variation.\footnote{The second corrects for zero point offsets in RV measurements, but the magnitude of the correction for J1446 is smaller than 1$\sigma$.}    

For the high-contrast imaging data, we observed J1446 with Keck/NIRC2+pyWFS \citep{Bond2020} at $L'$-band ($\lambda_{\rm cen}=3.776\ \mu{\rm m}, \Delta\lambda=0.7\ \mu{\rm m}$) in 2023 August and 2024 January (PI: Charles Beichman, see Table~\ref{tab: obs log} for details). 
We kept the PSF core below the saturation limit so that we could use the stellar PSF as a PSF reference for forward-modeling and photometry.

\begin{deluxetable}{ccccc}
\label{tab: obs log}
    \tablecaption{Keck/NIRC2 Observing Logs}
    \tablehead{
    \colhead{date [UT]}  & \colhead{${t_{\rm exp}}^a$ [sec]} & \colhead{$N_{\rm exp}\ ^b$} & \colhead{FWHM [pix]} & \colhead{field rotation}
    } 
\startdata
      2023 Aug 15 & 28 & 52 & 7.4 & 20\fdg0 \\
      2024 Jan 22 & 30 & 72 & 7.3 & 43\fdg9 
\enddata
    \tablecomments{a: The product of single exposure and the number of coadds. b: After removing poor-AO exposures.}
\end{deluxetable}

\subsection{Post-processing of Keck/NIRC2 data}

We used {\tt pyklip} \citep{pyklip} to subtract stellar PSF by generating a reference PSF through the Karhunen-Lo\`eve Image Processing algorithm \citep[KLIP;][]{Soummer2012}.
As the companion (see Section~\ref{sec: Results}) is located very close to the central star, we modified the {\tt pyklip} post-processing parameters to retrieve the companion's signal while sufficiently suppress residual speckles. After cropping the NIRC2 FoV into 620 by 620 pixels ($\sim$~6\arcsec by 6 \arcsec), we adopted annuli = 7 for the first epoch and 3 for the second epoch, and subsections = 4, which divide the cropped FoV into several partial annulus areas where localized reference PSFs are generated for angular differential imaging \citep[ADI;][]{Marois2006}. We finally adopted a Karhunen-Lo\`eve (KL) mode of 3 for the first epoch data and 10 for the second epoch data.
Note that pyWFS data needs to be corrected by an additional $0\fdg118\pm0\fdg006$ angle offset counterclockwise relative to Shack-Hartmann (SH) WFS data, which was measured by comparing NIRC2+pyWFS data of M92 with distortion-calibrated HST data (Walker et al. submitted). We have corrected this offset for the position angle measurements of the point source in Section~\ref{sec: Results}\footnote{We replaced $\beta=0\fdg262$ with $\beta=0\fdg380$, where $\beta$ is the NIRC2 orientation angle \citep[][see also \href{https://github.com/jluastro/nirc2_distortion/wiki}{NIRC2 distortion wiki} for the latest $\beta$ information with SH WFS]{Yelda2010,Service2016}.}. Regarding the pixel scale, we used the same value as \cite{Service2016} ($9\farcs971\pm0\farcs004$~mas/pixel) because Walker et al. derived a consistent scale of the pyWFS data within 3$\sigma$.

\section{Results} \label{sec: Results}

Table~\ref{tab: RV} summarizes the RV measurements for J1446. In contrast to the IRD-SSP targets presented in \cite{Uyama2023} that show a linear RV trend, the IRD RV monitoring of this system shows a curvature of the RV perturbations, which is helpful to improve orbital fitting of the companion (see Section~\ref{sec: Discussions}).

\begin{table}[]
\centering
\caption{RV Measurements of J1446}
\label{tab: RV}
\begin{tabular}{ccc}
\hline
JD & RV [m/s] & RV Error [m/s] \\ \hline
2458628.85673 & -5104.39 & 6.02 \\
2458628.86091 & -5112.21 & 6.01 \\
2458628.86495 & -5103.84 & 6.38 \\
2458628.86897 & -5093.55 & 6.50 \\
2458652.89532 & -5157.22 & 4.03 \\
2458652.90335 & -5159.80 & 4.61 \\
2459005.85503 & -5666.64 & 6.63 \\
2459005.86236 & -5667.64 & 5.52 \\
2459304.98868 & -5942.34 & 3.42 \\
2459324.01389 & -5949.02 & 3.28 \\
2459597.09216 & -6033.25 & 3.65 \\
2460419.85878 & -5581.20 & 3.97 \\
2460723.12214 & -5286.86 & 4.82 \\
2460745.96469 & -5235.99 & 4.60 \\
\hline
\end{tabular}
\end{table}

\begin{figure*}
\begin{tabular}{cc}
    \begin{minipage}{0.4\hsize}
    \centering
    \includegraphics[width=\linewidth]{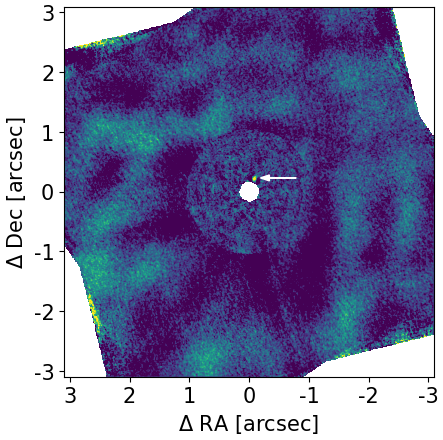}
    \caption{Post-processed NIRC2 image of the J1446 data taken in 2023 August (see text for the post-processing parameters). The white arrow indicates the location of the new companion. The central star is masked by the post-processing algorithms.}
    \label{fig: ADI}
    \end{minipage}
    \begin{minipage}{0.6\hsize}
    \centering
    \includegraphics[width=\linewidth]{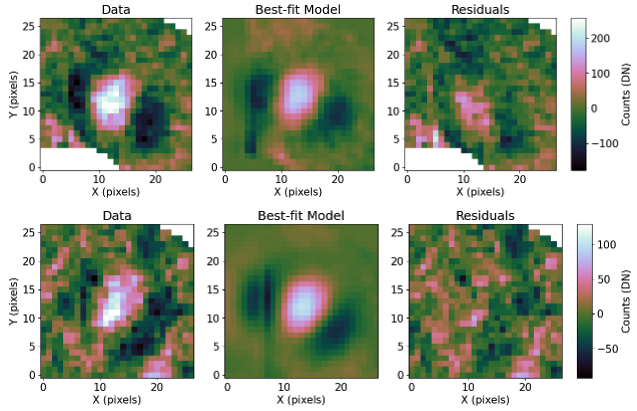}
    \caption{Left: Original post-processed image (top:  the 2023 August data with KL=3, bottom: the 2024 January data with KL=10) at the location of a point-like source. Middle: The best-fit forward-modeled PSF. Right: Residual map after subtracting the forward-modeled PSF from the post-processed image.}
    \label{fig: PSF fitting}
    \end{minipage}
\end{tabular}
\end{figure*}

Figure~\ref{fig: ADI} shows the post-processed result of the 2023 August Keck/NIRC2 data and a point-like source is detected close to the central star. We did not detect any other sources in the NIRC2 data.
We detected this source in both epoch data sets and we then used the combined stellar PSF as a reference PSF for forward modeling and MCMC PSF fitting using the {\tt pyklip} modules \citep[with the Matern 3/2 kernel;][]{Wang2016,Pueyo2016,emcee} to extract photometry and astrometry of this source. Figure~\ref{fig: PSF fitting} shows the PSF fitting results of the point-like source for both epochs, which is well fitted with the reference PSF scaled by a contrast. 
The extracted photometry, signal-to-noise ratio (SNR), and relative astrometry are summarized in Table~\ref{tab: obs result}. Although the second epoch obtained more total exposure time and field rotation for ADI reduction, the derived SNR is higher for the first epoch because the companion candidate is located closer to the central star in the second epoch.

\begin{table*}[]
    \centering
    \caption{Extracted Contrast, Photometry, and Astrometry of J1446B}
    \begin{tabular}{cccccc}
         MJD & Contrast & $L'$-band mag & SNR & Separation [mas] & PA   \\ \hline
         60171.2 & $3.6\times10^{-3} \pm 0.2\times10^{-3}$ & $15.22\pm0.07$ & 15.9 & $231.3 \pm 1.4$ & $338\fdg0\pm0\fdg3$ \\ 
         60331.6 & $5.4\times10^{-3} \pm 0.5\times10^{-3}$ & $14.78\pm0.11$ & 10.1 & $199.3 \pm 5.0$ & $339\fdg5 \pm 0\fdg6$ 
    \end{tabular}
    \label{tab: obs result}
\end{table*}

\begin{figure}
    \centering
    \includegraphics[width=\linewidth]{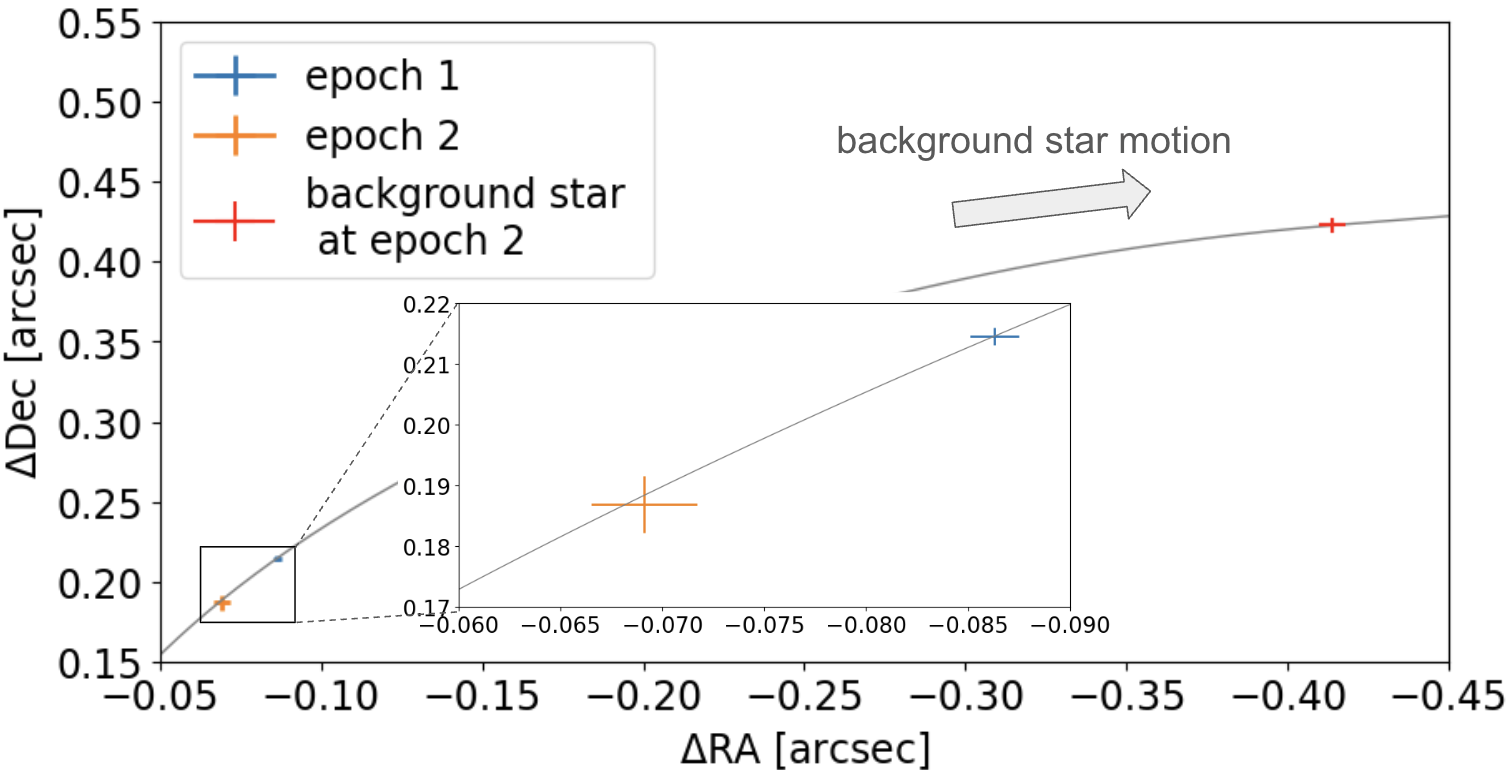}
    \caption{Common proper motion test of the companion candidate around J1446. The blue and orange crosses indicate the relative location to the central star ($\Delta{\rm RA}$ for x-axis and $\Delta{\rm Dec}$ for y-axis, respectively). The gray solid line indicates a trajectory of a zero proper-motion background star and the red cross indicates an expected position of the background-star case in the second epoch. We insert the zoomed-in panel for the relative astrometry of the first and second epochs, whose error-bars are small in the original panel. }
    \label{fig: cpm test}
\end{figure}

We finally conducted a common proper motion test to rule out the background star scenario, and Figure~\ref{fig: cpm test} demonstrates that the companion candidate is gravitationally bound to J1446.

\section{Discussions} \label{sec: Discussions}
As our proper motion test revealed that the source we newly detected is a gravitationally-bound companion to J1446, we then use photometric and astrometric information to characterize its physical parameters.

\subsection{Photometry} \label{sec: Photometry}
\begin{figure*}
    \centering
    \includegraphics[width=0.85\linewidth]{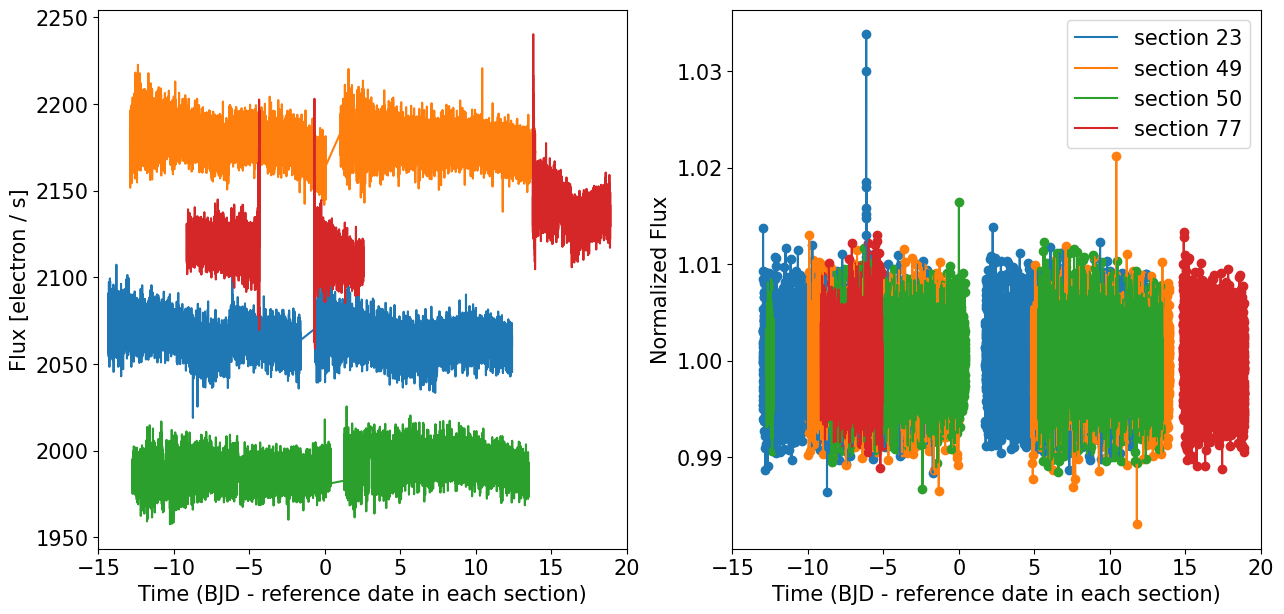}
    \caption{TESS lightcurves of J1446 ranging four sectors. Left: Comparisons of the measured flux variations in the four sectors, where we adopted median time of each section as a reference date. Right: The same comparison as the left panel except for the y-axis with normalized fluxes.}
    \label{fig: TESS}
\end{figure*}

The second-epoch contrast is brighter than the first-epoch contrast by 33\% ($\approx3.1\sigma$). We first looked into potential variability in the stellar flux. TESS monitored J1446 in several sectors and we downloaded the lightcurves. Figure~\ref{fig: TESS} shows the Science Processing Operations Center lightcurves, processed by {\tt lightkurve} \citep[][DOI of the TESS data; \href{http://dx.doi.org/10.17909/5zrm-fq94}{10.17909/5zrm-fq94}]{lightkurve}, and apparently J1446 shows short-term (hourly-daily) variations by up to $\sim$1\%, except for a flare event ($\sim3\%$) happened in sector~23, and longer-term (monthly-yearly) variations by $\sim3.5\%$ across these sectors, which can be systematics among different sectors. 
Although the variations at optical and NIR wavelengths may be caused by different mechanisms \citep[cf,][]{Petrucci2024}, it is unlikely that the NIR variability is larger than the optical variability by an order of magnitude. Therefore the variation in contrast should be mainly attributed to that of the companion. To convert the contrast into apparent flux, we adopted the mean contrast value of the two epochs ($4.5\times10^{-3}$) and the apparent magnitude at $L'$-band corresponds to $15.02\pm0.19$~mag \citep[with {\tt synphot};][we derived $L'$-band magnitude of the central star as $9.11\pm0.02$]{synphot}. Note that the error bar of the adopted apparent magnitude includes the scatter between the two epochs as we expect variability.
A precise age estimation is technically difficult for such an individual nearby M~dwarf and we assumed 1--10~Gyr for the age range. With COND03 evolutionary model \citep{Baraffe2003}, the derived apparent magnitude is consistent with a $\sim20\ M_{\rm Jup}$ object at 1~Gyr and a $\sim60\ M_{\rm Jup}$ object at 10~Gyr. In either case the $L'$-band flux is consistent with a mid/late T-dwarf regime, less than 1000~K with the COND03 model.

In the 2023 August run, we obtained ($\lambda_{\rm cen}=2.1686\ \mu{\rm m}, \Delta\lambda=0.0326\ \mu{\rm m}$) short exposures of J1446 with the Br$\gamma$ filter (20~seconds$\times$5~frames) as a screening process before the deep $L'$-band observations, and we utilized other short Br$\gamma$ data sets of LSPM~J0015+1333 \citep[used in][]{Kuzuhara2024} and LSPM~J2212+0833 (the results of the continued program is in preparations) taken on the same date, where we did not see bright companions, as PSF reference stars and conducted reference-star differential imaging reduction. As a result we did not confirm J1446B's signal but calculated the contrast limit at $0\farcs2$ as $1.7\times10^{-2}$. Although this contrast limit does not sufficiently constrain details of its characteristics in a color-magnitude diagram, the non-detection at the Br$\gamma$ filter favors a substellar-mass object.

\subsection{Orbital Fitting} \label{sec: Orbital Fitting}

We could not use Hipparcos-Gaia proper motion acceleration \citep{Brandt2018} to constrain the orbit and estimate the dynamical mass because Hipparcos did not record J1446's astrometry. However, J1446 is present in Gaia DR3 as an accelerating, non-single-star fit \citep{GaiaDR3,GaiaDR3_NSS}.  Specifically, it has a seven parameter fit, adding an acceleration in both R.A.~and Decl.~in to the two positions, two proper motions, and parallax of a standard five-parameter fit.  We also have a measured RV time series of the primary star, which is complementary to astrometry for estimating the dynamical mass combined with direct imaging. In this section we present two orbital fits.  First, we perform a fit only using RVs of the primary star and relative astrometry from direct imaging (DI).  Second, we add the astrometric accelerations reported by Gaia DR3. 

Our orbital fitting provides a constraint on the dynamical mass of a substellar-mass object around an M~dwarf. Prior to this work, the dynamical masses of only a few BD companions around nearby M~dwarfs were constrained; \cite{Dupuy2019} constrained the dynamical mass of a directly-imaged BD companion WISE~J072003.20-084651.2B (Scholz’s star) combining adaptive optics imaging and ground-based astrometry. \cite{Brandt2020} constrained Gliese~229B's mass combining radial velocity, Hipparcos/Gaia absolute astrometry, and direct imaging results, and \cite{Thompson2025} constrained the dynamical mass of each BD \citep[Gliese~229Bab;][]{Xuan2024}. 

\subsubsection{RV+DI orbital fitting} \label{sec: RV+DI orbital fitting}
With the IRD RV data and NIRC2 high-contrast imaging, we conducted orbital fitting using {\tt orvara} \citep{Brandt2021}. Combining RV and direct imaging relative astrometry can sufficiently solve degeneracies of orbital parameters such as the dynamical mass of the companion and the inclination. For the priors we adopted $M_{\rm pri}=0.2\pm0.1 M_\odot$, which covers an expected mass from the previous characterizations \citep[$\sim0.15-0.16\ M_\odot$][]{} and typical mass of a mid-M star from the CARMENES survey \citep[$\sim0.2-0.3\ M_\odot$][]{Cifuentes2020}, and referred to the main source catalog of Gaia DR3 \citep{GaiaDR3} for the parallax ($58.63\pm0.11~{\rm mas}$).

Table~\ref{tab: orbital parameters} summarizes the derived orbital parameters, and the derived dynamical mass for J1446B is $M_{\rm sec}={62.1}_{-1.5}^{+1.5}\ M_{\rm Jup}$.
However, our fitting suggests the central star is heavier than the typical mass of a mid-M star \citep{Cifuentes2020} by a factor of $\sim2$. 
We first tested another orbital fitting tool used in \cite{Feng2023} and \cite{Xiao2024} and obtained consistent posteriors with the {\tt orvara} fitting. We also looked for signature of an unseen companion using the TESS light curves and the IRD spectra (see Appendix~\ref{sec: CCF analysis}) but did not find evidence of binarity.

We then tested {\tt orvara} fitting with stricter priors of 1) $M_{\rm pri}=0.15\pm0.05 M_\odot$ for the mass with 50,000 steps (baseline: 10,000 steps in the original {\tt orvara} fitting), and 2) $M_{\rm pri}=0.155\pm0.001 M_\odot$ and $58.633\pm0.001$~mas for the parallax with 100,000 steps. For case 1, the posterior of the primary mass is $0.377\pm0.003 M_\odot$, which shows the same trend as the original {\tt orvara} fitting, favoring a significantly higher mass than the prior. The posterior of the secondary is close ($60.3\pm1.4 M_{\rm Jup}$) to the original fitting result. Case 2 did not converge as well as the original fitting despite the longer run; the ensemble chains failed to mix, remaining nonstationary with long-term drifts in several parameters, posteriors remained multi-modal, and the posterior-predicted orbits showed significant and systematic offsets from the input RV data. Therefore, the cause of the observed stellar mass discrepancy appears to be unrelated to the specific orbit-fit settings.

RV and DI measure the system mass through the semimajor axis (via DI) and the period (primarily via RV in this case), together with Kepler's Third Law.  The resulting system mass is very sensitive to systematics and underestimated uncertainties in any of these parameters, particularly since our data cover less than half the orbit (with data extending back to 2019).  The inferred mass of the brown dwarf is more robust because RV probes the acceleration of the primary star in response to the tug of its companion.  Gaia DR3 extends our data several years further into the past, with an acceleration measured around 2016.  In the following section, we incorporate these data into our orbital fit.


\begin{table*}
\centering
\caption{Orbital Parameters of the J1446 System}
\label{tab: orbital parameters}
\begin{tabular}{ccccc} 
parameter & \multicolumn{2}{c}{RV+DI} & \multicolumn{2}{c}{RV+DI+Gaia acceleration} \\ \hline
 & priors$^a$ & posteriors & priors$^a$ & posteriors \\
\hline\hline
Jitter [m/s]     & 1--100  &  ${7.9}_{-2.5}^{+3.3}$ & 1--100 & ${8.4}_{-2.5}^{+3.3}$ \\
parallax [mas]  &  $58.63\pm0.11$ &  $58.63\pm 0.13$ & $58.303\pm0.018$ & $58.303\pm0.018$  \\
$M_{\rm pri}$ [$M_\odot$]  & $0.2\pm0.1$ &      ${0.457}_{-0.034}^{+0.036}$ & $0.2\pm0.1$ & ${0.151}_{-0.021}^{+0.023}$ \\
$M_{\rm sec}$ [$M_{\rm Jup}$]    & \dots  &   $62.1\pm1.5$ & \dots & ${59.8}_{-1.4}^{+1.5}$ \\
a [AU]        & \dots  & $4.66\pm0.16$ & \dots & $4.30\pm0.06$ \\
inclination [deg] & \dots &      $85.7\pm2.0$ & \dots & $ 85.3 \pm 1.5 $ \\
ascending node [deg] & \dots &  $154.8\pm1.7$ & \dots & $156.2\pm0.8$ \\
mean longitude [deg] & \dots &      ${217}_{-16}^{+15}$ & \dots & ${303.8}_{-8.4}^{+7.9}$ \\
period [yrs]  & \dots &  ${13.98}_{-0.67}^{+0.73}$ & \dots & ${19.52}_{-0.76}^{+0.78}$ \\
argument of periastron [deg]   & \dots &    ${89}_{-19}^{+17}$ & \dots & ${109.0}_{-3.3}^{+2.4}$ \\
eccentricity   & \dots  &    ${0.105}_{-0.010}^{+0.012}$ & \dots & ${0.122}_{-0.015}^{+0.014}$ \\
T0 [JD]       & \dots & ${2458496}_{-212}^{+201}$ & \dots &${2458472}_{-95}^{+71}$ \\
mass ratio    & \dots &  ${0.130}_{-0.008}^{+0.009}$ & \dots & ${0.379}_{-0.054}^{+0.069}$
\end{tabular}
    \tablecomments{a: We show only specified priors in the priors column, and other priors follow default settings in {\tt orvara}. The minimum and maximum values of Jitter indicate the lower and upper bounds for the log-uniform prior. The parallax and primary mass assumed Gaussian priors. }
\end{table*}

\subsubsection{Incorporating Gaia proper motion acceleration into the RV+DI fitting} \label{sec: RV+DI+Gaia fitting}

Gaia DR3's accelerating fit offers a powerful source of additional data to constrain the system's orbit and measure the components' masses. The measured acceleration is  
$1.863 \pm 0.045$ mas\,yr$^{-1}$ in R.A.~and $-5.726 \pm 0.045$ mas\,yr$^{-1}$ in Decl., i.e, more than 100$\sigma$ significant.  This acceleration is measured between mid 2014 and mid 2017, much further back in time than our RV time series or our imaging. 
We implemented an orbital fit with Gaia DR3 accelerations by first generating a custom-built input catalog that can replace the HGCA \citep{Brandt2021-HGCA} for ingestion into {\tt orvara} \citep{Brandt2021}. This custom catalog includes the direct Gaia acceleration terms from the Gaia DR3 non-single-star astrometric solutions \citep{GaiaDR3_NSS}. These accelerations terms are modeled internally in {\tt orvara} using {\tt htof} \citep{htof}, which already supports accelerating fits. In fact, acceleration terms are already accounted for in the likelihood calculation for {\tt orvara} if they are given, but the properties of the underlying acceleration data have not been tested as thoroughly as the proper motions \citep{Brandt2021-HGCA}. We thus used {\tt orvara} with this custom catalog file to perform a joint orbit fit \citep[for details, see Section 4.2 of][]{An2025_RV+HGCA+Gaia}\footnote{Note that this analysis process is not yet available in a public branch of {\tt orvara}. An official implementation is planned for a future release to coincide with Gaia DR4, which is expected to publish a significantly larger number of astrometric acceleration solutions.}. 
\citet{An2025_RV+HGCA+Gaia} compared 11 Gaia non-single-star solutions and found that the Gaia acceleration uncertainties are underestimated (see Section 4.4 and Figure~11 in the reference). To address this, we inflated the uncertainties by a factor of two in the orbit fit.

\begin{figure*}
\begin{minipage}{0.5\hsize}
    \centering
    \includegraphics[width=0.95\linewidth]{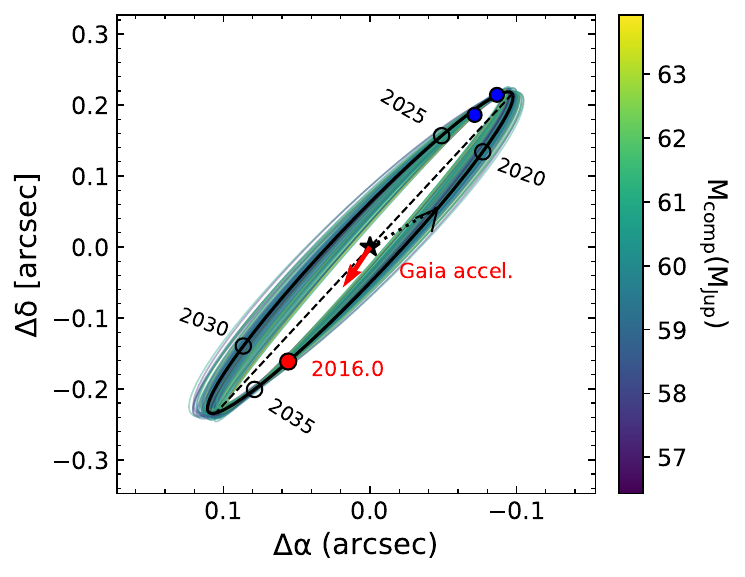}
\end{minipage}
\begin{minipage}{0.5\hsize}
    \centering
    \includegraphics[width=0.9\linewidth]{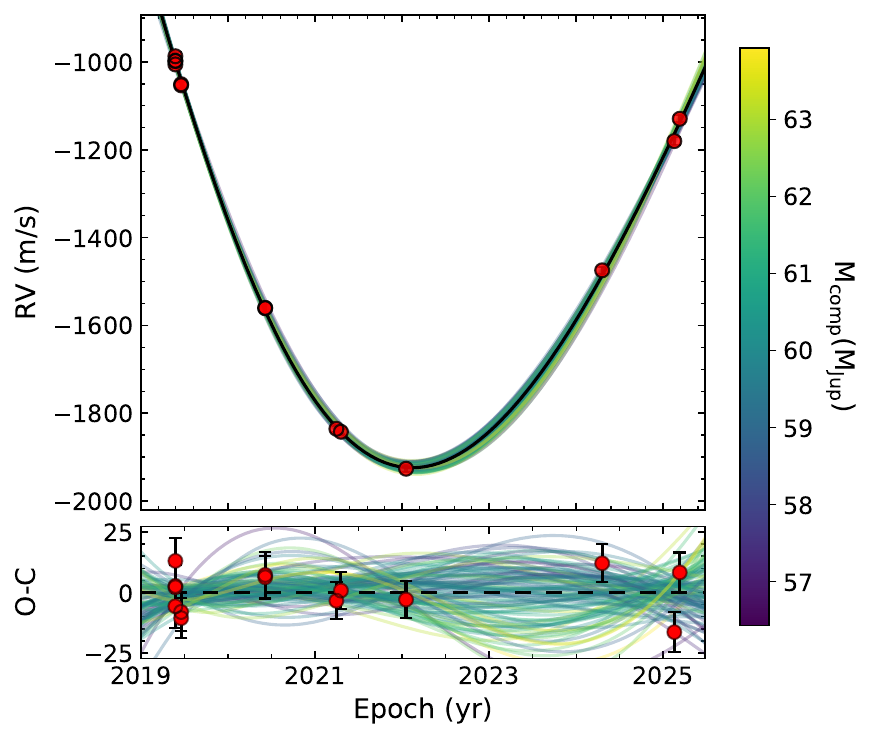}
\end{minipage}
    \caption{
    Left: Modeled projected orbits of J1446B from fitting the Gaia DR3 acceleration (a red arrow in left) and two measured Keck/NIRC2 relative astrometric measurements (blue dots in left), and Subaru/IRD RVs (red dots in right). The host star defines the origin and is indicated by a black star; one hundred random orbital draws are shown and color-coded by companion mass, with the best-fit orbit (black). The fitted offset at 2016.0 is plotted in red, and the direction of the measured Gaia DR3 acceleration is indicated by the red arrow. Right: Modeled RVs from a random sampling of orbits from the posterior are color-coded by companion mass with the best-fit orbit.
    }
    \label{fig: orvara fitting}
\end{figure*}

The Gaia DR3 acceleration vector points to the position of the brown dwarf at that time ( left panel of Figure \ref{fig: orvara fitting}, denoted in red arrow and red dot), offering an important but indirect constraint on the orbital period.  In this section we incorporate the Gaia DR3 acceleration into the orbital fitting; this is the first application of Gaia accelerations to the orbit analysis of a real companion.  We further note that the parallax in the non-single-star fit, $58.304 \pm 0.017$~mas, is more precise than and differs slightly from the value of $58.63 \pm 0.11$ mas in the five-parameter fit.  We adopt the value from the accelerating star fit here, which was shown to be more reliable in the context of binary stars \citep{Nagarajan+ElBadry_2024}.

Combining Gaia proper-motion acceleration with RV and DI data yields a primary mass of $0.151^{+0.023}_{-0.021} M_\odot$ and a companion mass of $59.8^{+1.5}_{-1.4} M_{\rm Jup}$, and returns a fully constrained set of orbital parameters (Table \ref{tab: orbital parameters}). We used the same initial mass prior in this orbit fit ($0.2\pm0.1 M_\odot$) as the RV+DI fitting (Section~\ref{sec: RV+DI orbital fitting}). This fit resolves the earlier primary-mass discrepancy and provides a robust dynamical mass constraint on J1446B. The comparison of the dynamical mass with the $L'$-band flux (Section~\ref{sec: Photometry}) indicates that the system is consistent with $\sim10$~Gyr. In case the system is younger than $\sim10$~Gyr, it would indicate that B is a binary consisting of lower-mass BDs like Gliese~229B \citep{Xuan2024}. 
Figure \ref{fig: orvara fitting} plots the relative orbits of J1446B on sky plane (left panel), color coded by fitted companion mass. The direction of the Gaia DR3 acceleration vector and the fitted position of J1446B are overplotted in a red arrow and a red dot, respectively. The right panel plots RV orbits (color coded by fitted companion mass) of J1446 overplotted with observed RVs (in red dots), where the best-fit orbits are denoted in black.


\begin{figure*}
    \centering
    \includegraphics[width=0.9\textwidth]{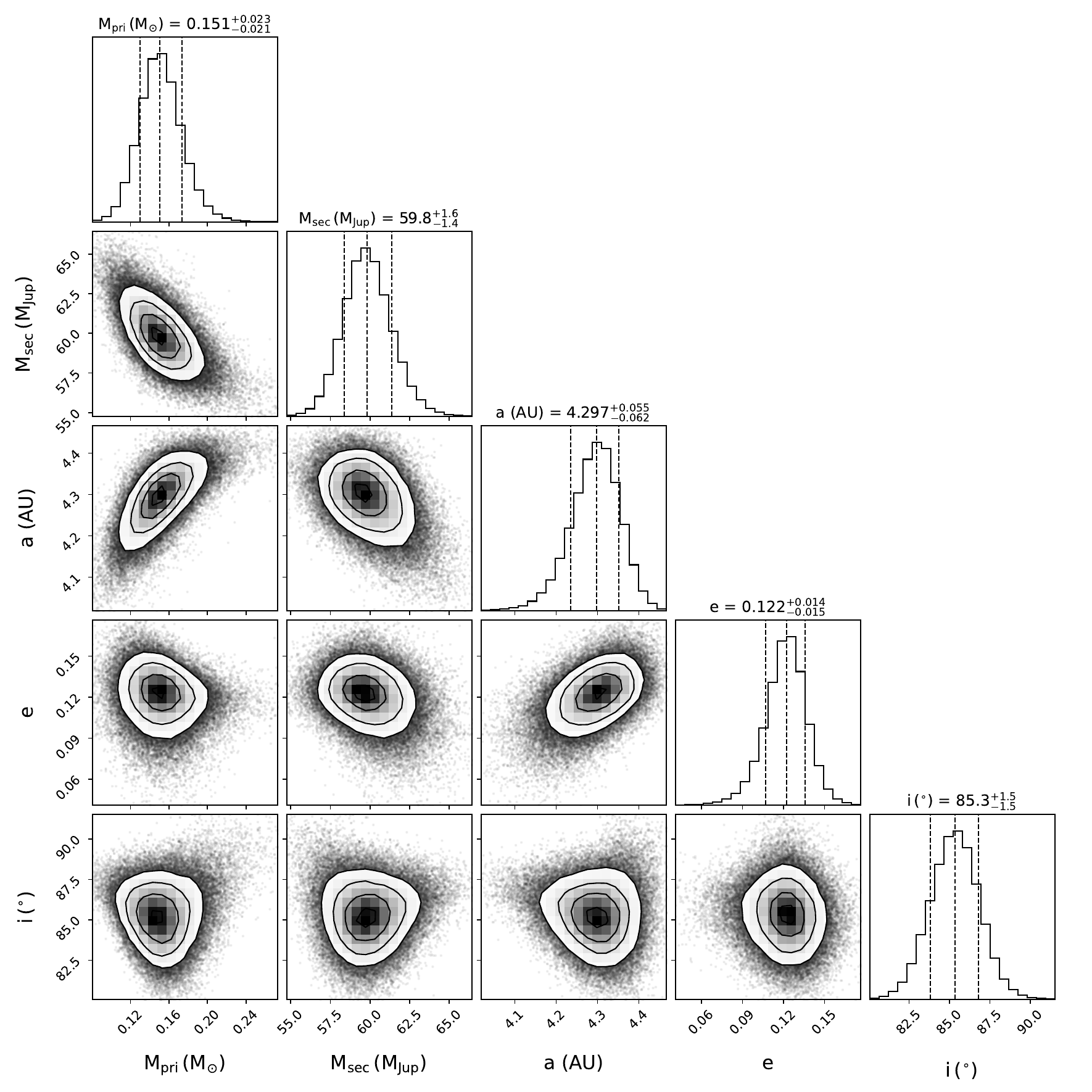}
    \caption{Corner plot of the orbital fitting result combining Subaru/IRD radial velocity, Keck/NIRC2 direct imaging, and Gaia proper motion acceleration. The orbital fit is described in Section \ref{sec: Orbital Fitting}.}
    \label{fig: corner plot}
\end{figure*}

The dynamical mass ratio ($q\approx0.38$) and semi-major axis of J1446B ($a\approx4.3$~au) mark the J1446 system a very unique system among nearby M~dwarfs \citep[cf,][]{Winters2019,Comez2023}. J1446B is the third directly-imaged BD companion around an M~dwarf with a semi-major axis $\leq5\ {\rm AU}$ \citep[LHS 2397AB and Scholz's star][]{Dupuy2009,Burgasser2015}. While the host stars of the known BD companions except J1446 have spectral types of M8-M9 and their mass ratios are $q\sim0.7$, suggesting binary formation. On the other hand, J1446B's mass ratio is about half of these BD companions. 
In Section~\ref{sec: Photometry} we mentioned that the $L'$-band contrast is variable, which is unlikely to be due to stellar variability. BD variability is common among L/T transition objects \citep{Radigan2014} for various possible factors such as clouds and temperature variability, hot spot, and aurorae \citep[e.g.][]{Apai2017,Eriksson2019,Biller2024}, but only a few late T-dwarfs have been reported to exhibit strong variabilities \citep[][]{Manjavacas2019,Miles2025}. Therefore J1446B is one of the intriguing late-T~dwarfs showing variability at $L'$-band for future atmospheric studies with the constrained dynamical mass.
With the astrometric and photometric data from our observations, detailed investigations of the J1446 system will provide a hint about the formation scenario of the J1446 system and update the statistical discussions on M-dwarf multiplicity \citep[cf,][]{Ma2014}.

We note that our orbital fitting suggests that J1446B is moving inwards as of 2025 and follow-up high-contrast imaging/spectroscopic observations will be technically more feasible around 2030 when the projected separation of J1446B becomes large (see the left panel of Figure~\ref{fig: orvara fitting}). Future Gaia data releases will provide Intermediate Astrometric Data like Hipparcos \citep[e.g.,][]{Nielsen2020} as well as astrometric solutions, and then orbital fitting will improve the constraints on these parameters.

\section{Conclusion} \label{sec: Conclusion}

We presented the Keck/NIRC2+pyWFS discovery of a new brown-dwarf companion around a nearby mid-M~dwarf J1446, which was suggested by RV variations from the Subaru/IRD-SSP survey. The $L'$-band flux is consistent with a cool brown dwarf and the contrast shows significant ($\sim30\%, 3.1\sigma$) variability that is unlikely to be due to stellar variability. Our orbital fitting combining the Subaru/IRD RV monitoring results, relative astrometry from the NIRC2 direct imaging data sets, and Gaia non-single-star catalog for acceleration indicates ${59.8}_{-1.4}^{+1.5}\ M_{\rm Jup}$ for J1446B's dynamical mass. Our work demonstrated that Gaia DR3's acceleration provides strong constraints on the orbital parameters of J1446B, for which our RV and DI observations covered less than half of the orbital period.
The derived mass ratio ($q\approx0.38$) and semi-major axis ($a\approx4.3~{\rm au}$) mark J1446B a very unique BD companion among the directly-imaged BD companions around nearby M~dwarfs. 
Future Gaia Data Releases, including Intermediate Astrometric Data and astrometric solutions, will be able to improve orbital fitting.
Follow-up high-contrast imaging or spectroscopy, which will be feasible in $\sim2030$, will provide important information about the atmosphere and variability of this new intriguing BD companion around an M~dwarf. Particularly future high-dispersion spectroscopy such as Keck/HiSPEC or TMT/MODHIS \citep{Konopacky2023} could measure RV of the companion, which can improve orbital fitting \citep[e.g.,][]{Hsu2024} as well as future Gaia Data releases.

\section*{Acknowledgments}

The authors would like to thank the anonymous referee for their constructive comments and suggestions to improve the quality of the paper.
We thank Guangyao Xiao for help in implementing fitting tool in \cite{Feng2023}.
M.K. and J.K. are supported by JSPS KAKENHI (grant No. 24K07108 for M.K. and 24K07086 for J.K). D.S., and T.T. acknowledge support by the BNSF program "VIHREN-2021" project No. KP-06-DV/5.
The development and operation of IRD were supported by JSPS KAKENHI Grant Nos. 18H05442, 15H02063, and 22000005, and Astrobilogy Center (ABC) of NINS. 
This work was supported in part by the NSF Graduate Research Fellowship (grant No. 2139433).

This research is based on data collected at the Subaru Telescope, which is operated by the National Astronomical Observatory of Japan. 
The part of data presented herein were obtained at the W. M. Keck Observatory, which is operated as a scientific partnership among the California Institute of Technology, the University of California and the National Aeronautics and Space Administration. The Observatory was made possible by the generous financial support of the W. M. Keck Foundation.
The authors wish to recognize and acknowledge the very significant cultural role and reverence that the summit of Maunakea has always had within the indigenous Hawaiian community. We are most fortunate to have the opportunity to conduct observations from this mountain.
Data analysis was in part carried out on the Multi-wavelength Data Analysis System operated by the Astronomy Data Center (ADC), National Astronomical Observatory of Japan.
This work has made use of data from the European Space Agency (ESA) mission
{\it Gaia} (\url{https://www.cosmos.esa.int/gaia}), processed by the {\it Gaia} Data Processing and Analysis Consortium (DPAC,
\url{https://www.cosmos.esa.int/web/gaia/dpac/consortium}). Funding for the DPAC has been provided by national institutions, in particular the institutions participating in the {\it Gaia} Multilateral Agreement. 
Funding for the TESS mission is provided by NASA's Science Mission Directorate. We acknowledge the use of public TESS data from pipelines at the TESS Science Office and at the TESS Science Processing Operations Center (SPOC). Resources supporting this work were provided by the NASA High-End Computing (HEC) Program through the NASA Advanced Supercomputing (NAS) Division at Ames Research Center for the production of the SPOC data products. This paper includes data collected by the TESS mission that are publicly available from the Mikulski Archive for Space Telescopes (MAST).
This research made use of Lightkurve, a Python package for Kepler and TESS data analysis (Lightkurve Collaboration, 2018).
This publication makes use of VOSA, developed under the Spanish Virtual Observatory (\url{https://svo.cab.inta-csic.es}) project funded by MCIN/AEI/10.13039/501100011033/ through grant PID2020-112949GB-I00.
This research has made use of NASA's Astrophysics Data System Bibliographic Services.
This research has made use of the SIMBAD database, operated at CDS, Strasbourg, France.

\appendix
\section{Additional unseen companion?} \label{sec: CCF analysis}

Although our orbital fitting incorporating radial velocity, direct imaging, and Gaia proper motion acceleration resulted in the consistent mass for the primary star with the SED, we discuss a potential scenario that could explain the discrepancy between RV+DI orbital fitting with the SED (Section~\ref{sec: RV+DI orbital fitting}); the central star is a tight binary and the RV variations are caused by the other unseen companion.

\begin{figure*}
\centering
    \includegraphics[width=0.9\textwidth]{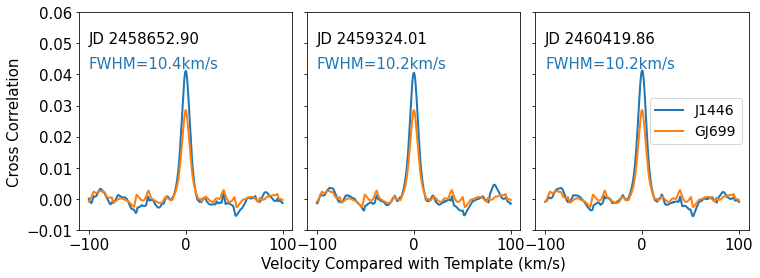}
    \caption{CCFs between the IRD spectra of J1446 and a reference signle star GJ~699 over three epochs, showing single-peaked profiles and similarity between these two stars.}
    \label{fig: CCF}
\end{figure*}

While TESS data shows some variabilities in its light curve (Figure~\ref{fig: TESS}), M dwarfs often show variability \citep[e.g.,][]{Kar2024}. Our analysis on the TESS light curves suggest that the central star is unlikely an eclipsing binary. We investigated the IRD spectra in detail if we see evidence of binarity such as double-line features or broadened absorption lines \citep[e.g.,][]{Merle2017}.
Following \cite{Mori2022}, we computed cross correlation functions (CCFs) between J1446’s spectra and an IRD spectrum of a well-known single star GJ~699 (Barnard’s Star) as a reference. Figure~\ref{fig: CCF} shows CCFs at three epochs when each single SNR of J1446 was sufficient to calculate a CCF, and J1446’s profiles in these epochs are single-peaked. If J1446 is actually a binary, its orbit should be almost face-on that IRD could not resolve an RV shift between the two stars ($\lesssim$ 4~km/s), which would suggest a misaligned orbit from J1446B by up to $\sim90^\circ$.

\bibliography{library}                                    
\end{document}